# Periodicity doubling cascades: direct observation in ferroelastic materials

*Arnoud S. Everhardt[1#], Silvia Damerio*[1], Jacob A. Zorn[2], Silang Zhou[1], Neus Domingo[3], Gustau Catalan[3,4], Ekhard K. H. Salje[5], Long-Qing Chen[2] and Beatriz Noheda*[1,6]*

[1] Zernike Institute for Advanced Materials, University of Groningen, Nijenborgh 4, 9747 AG Gronigen, The Netherlands

[2] Department of Materials Science and Engineering, The Pennsylvania State University, University Park, Pennsylvania 16802, United States of America

[3] Catalan Institute of Nanoscience and Nanotechnology (ICN2), CSIC, Barcelona Institute of Science and Technology Campus, Universitat Autonoma de Barcelona, Bellaterra, 08193 Barcelona, Spain

[4] ICREA, 08193 Barcelona, Spain

[5] University of Cambridge, UK

[6] Groningen Cognitive Systems and Materials Centre (CogniGron), University of Groningen, The Netherlands

ABSTRACT

Very sensitive responses to external forces are found near phase transitions. However, phase transition dynamics and pre-equilibrium phenomena are difficult to detect and control. We have directly observed that the equilibrium domain structure following a phase transition in BaTiO$_3$, a ferroelectric and ferroelastic material, is attained by halving of the domain periodicity, sequentially and multiple times. The process is reversible, displaying periodicity doubling as temperature is increased. This observation is backed theoretically and can explain the fingerprints of domain period multiplicity observed in other systems, strongly suggesting this as a general model for pattern formation during phase transitions in ferroelastic materials.

TEXT

(INTRODUCTION)

The current interest in adaptable electronics calls for new paradigms of material systems with multiple metastable states. Functional materials with modulated phases[1] bring interesting possibilities in this direction, as long as we are able to tune the stability of these available states. Ferroic materials are good prospective candidates because the modulation can be controlled by external magnetic, electric or stress fields. In magnetic materials, the presence of competing interaction between spins can lead to commensurate or incommensurate modulated structures, such as in the well-known case of the axial next-nearest-neighbor Ising (ANNNI) model.[2] In ferroelectric materials, interesting modulations in form of domain patterns, involve not only domains with alternating up and down polarization but also vortices[3] and ferroelectric skyrmions[4] as recently reported. Ferroelectrics are often also ferroelastic[5] and modulations of the strain state also occur in these materials. So far, measurement techniques such as X-Ray Diffraction (XRD)[6,7], Piezoelectric Force Microscopy (PFM) or Transmission Electron Microscopy (TEM) have been used to follow electric-field induced ferroelectric domain switching.[8–11] In contrast, microscopy studies of the temperature-driven annihilation of ferroelastic a/c nanodomains have received much less attention[12,13] and it is only recently that the experimental developments have allowed the in-situ study of domain dynamics.[14–19]

From a computer simulation point of view, the phase-field method has been extensively employed to model domain evolution and domain structures in a variety of ferroelectric systems.[20-22] While analytical thermodynamic theories are able to predict relative stability of single domain states or simple multidomain states with pre-assumed domain wall orientations[23], phase-field methods predict the stable domain states and the spatial scale under a given external constraint.

When ferroelastic materials are grown in thin film form on a suitable crystalline substrate, they can be subjected to epitaxial strain, which typically relaxes by the formation of ferroelastic domains.[24,25] For example, in a ferroelastic tetragonal system on a cubic substrate, 90° domains are expected to form to relieve the stress. Depending on the sign of the strain (compressive or tensile) imposed on the film by the substrate, either so-called a/c domains (with the long axis alternating in-plane and out-of-plane) or a/b domains (long axis fully in-plane) are preferred.[26] In the absence of dislocations or other defects, ferroelastic domains in epitaxial films are expected to alternate periodically.[24–26]

The periodicity of this modulation, or the domain width ($w$), is determined by the competition between the elastic energy in the domains and the formation energy of domain walls and is a function of the thin film thickness ($d$). For the cases when $d > w$, there is a square root dependence[24] $w = \beta\, d^{1/2}$, that holds for both epitaxial[27] and freestanding[28] layers. It is, in fact, a particular case of the Kittel's law that also applies to ferromagnetic and ferroelectric, non-ferroelastic, systems with 180° domains.[27,29] In the ferroelastic case, geometrical effects are also important and the phenomenon of discretization of domain widths, with minimum sizes determined by the need of lateral lattice coherence at the domain wall, have also been shown.[30,31] In addition, a careful look at the literature reveals that, often, the different domain sizes present in a particular film are related to each other (see Figure S1 in supplementary information).[32-36] Why this is the case is an open question. Recently, an original approach has been put forward, in which non-equilibrium ferroic domain structures in perovskite oxides are rationalized with respect to surface folding, wrinkling, and relaxation, following a hydrodynamics-like approach.[37] As both misfit strain and domain wall energies change with temperature, the equilibrium wall-to-wall distance is also temperature dependent and changes the domain pattern via coarsening. Thus, the question arises how does the system evolves towards (global or local) equilibrium. This is important because understanding the dynamics of domain formation will provide access to, and control on, the different available states.

In the present work, we report the direct observation of ferroelastic/ferroelectric domain evolution by sequential periodicity halving/doubling on $BaTiO_3$ thin films, using temperature-dependent piezoreponse-force microscopy (PFM). The mechanism observed agrees with expectations based on basic principles, but rarely discussed explicitly, of new domains forming at equidistant points between existing ones, where the stress is minimum. It is challenging to find an experimental system in which this process can be observed neatly and live.[38] Moreover, we present the results of phase field simulations that support the experimental observation of periodicity doubling in BaTiO$_3$ thin films. This phenomenon can explain the discretization of domain sizes that we have detected in various systems[32-36], suggesting that this is general for ferroelastic systems.

(EXPERIMENTAL DETAILS AND THEORETICAL METHODS)

The experiments have been performed on BaTiO$_3$ films grown under low epitaxial strain on SrRuO$_3$-buffered NdScO$_3$ substrates, as described in ref.[39]. The low-strain condition flattens the energy landscape such that different ferroelectric domain configurations can be accessed within a moderate temperature range.[40] In particular, the films display a paraelectric-ferroelectric phase transition at 130°C and a second transition at about 50°C. This second phase transition takes place in between two quite complex phases (details reported somewhere else[41]) but, for the sake of the present results, it can simply be described as a transition from a ferroelectric/ferroelastic high-temperature, pseudo-tetragonal a/c domain structure to a ferroelectric/ferroelastic low temperature, so-called ca$_1$/ca$_2$, monoclinic domain structure[41]. Since the monoclinic distortion and out-of-plane polarization components are rather small, this phase is very close to an orthorhombic phase (the two different in-plane lattice parameters are due to the in-plane anisotropy of the NdScO$_3$ substrate), with an in-plane 90° a/b domain configuration. Thus, for simplicity we treat the room temperature state as an in-plane a/b pseudo-tetragonal domain configuration.

Unlike in other ferroics, the transformation from a/c to a/b domains in these films is slow enough to be followed with the PFM technique. The experiments have been performed using Dual AC Resonance Tracking mode (DART) in an Asylum AFM system. This mode provides a natural signal amplification mechanism by following the cantilever mechanical contact resonance,. Since for these samples the polarization is mainly in-plane, we performed Lateral – PFM measurements (LPFM) and used the lateral deflection signal to extract the corresponding lateral cantilever contact resonance, which is sensitive to the in-plane ferroelectric polarization perpendicular to cantilever axis (DART - LPFM). After collecting PFM images of the films at room temperature, the samples were heated to the paraelectric phase at 200°C for 5 minutes and then cooled down in steps of 10ºC. Measurements were performed at each temperature between 70ºC (about 20ºC degrees above the nominal transition temperature), and 30°C. The same measurement procedure has been followed to study the domains evolution upon heating: the temperature was increased from 30ºC to 200°C in steps of 10ºC and PFM images were collected at each step between 30ºC and 70°C. The analysis of the PFM images was performed by means of the Gwyddion software and domains periodicities have been extrapolated from the Fast Fourier Transform (FFT) of the Lateral PFM Images.

The phase field method was utilized to model BaTiO$_3$ thin films under a relatively small anisotropic epitaxial strain. The bulk free energy density of the system was described by an eighth order Landau-Devonshire polynomial where the free energy coefficients were provided by Wang *et al*.[42] Both the electrostatic and elastic free energy contributions were also considered, and the methods provided by Li *et al*. were used to solve the resulting mechanical and electrostatic equilibrium equations.[22,43] The domain wall energy is assumed to be isotropic and introduced via a Ginzburg gradient energy.[23,44] We vary the temperature to obtain the domain pattern evolution during cooling or heating between 25°C and 100°C. Both 3D and quasi-2D simulations were utilized to understand the domain evolution behavior of the BaTiO$_3$ films.

(RESULTS and DISCUSSION)

An LPFM image showing the room temperature domain configuration prior to thermal cycling is shown in Figure 1a). This image shows the periodic a/b domains with domain walls parallel to the [110] direction (diagonal of the image). The periodicity of the initial a/b domains is $w_o$=(110 ± 10) nm (with a-domains of width $w_a$= 60 nm and b-domains $w_b$=50 nm). In some samples, the analysis of LPFM images by FFT together with the characterization via XRD also reveal the presence at room temperature periodicity peaks corresponding to the a/c domain configuration (see Supplementary Information S3). The coexistence of the two structures is not surprising given the marked first-order character of the transition. This enhances the complexity of the structure and is responsible for the relatively weak contrast and the need for resonance tracking to observe the domains in the PFM images.

Subsequently, the films were heated to 200 °C and measured during cooling, as presented in Figures 1b-f). During the cool-down process, the a/c structure arises below the para-ferro transition. The a/c domain structure is characterized by domain walls quasi-parallel to the [010] direction. At a temperature of 70 °C, the second phase transition starts and a single narrow b-domain can be found inside the a/c matrix (Figure 1b). In the amplitude LPFM images, b-domains are observed as dark lines since their in-plane component is parallel to the cantilever axis and, under this configuration, the LPFM is only sensitive to the in-plane ferroelectric polarization components that are perpendicular to the cantilever axis. Additional b-domains appear in Figure 1c) at 60°C. The domain sizes are measured to be (500 ± 50) nm, thus larger than the domain sizes in Figure 1a). Further cooling to 50°C (Figure 1d) and 40 °C (Figure 1e) shows some dark areas appearing in the middle of the a-domains. Further cooling down to 30°C (Figure 1f) reveals that those dark lines develop into new b-domains, reducing the domain sizes to (250 ± 20) nm, which is half of its value at 50°C and approximately double of the initial room temperature value of Figure 1a). The initial room temperature state is recovered upon cooling only after leaving the system to equilibrate for several hours. Thus, two sequential periodicity halving events have been observed during the transition from the a/c to the a/b domain configuration on cooling.

A similar phase transition mechanism, albeit shifted in temperature due to thermal hysteresis, is observed when heating the samples from the a/b to the a/c domain configuration, as shown in Figure 2. The a/b domain periodicity at 30°C is about $w_o$ = 100 nm. At 55°C, the domain walls start to rearrange and at 60°C, domains with $w = 2\,w_o$ are visible, as indicated by the green arrows in Figure 2c. With increasing the temperature up to 65°C (Figure 2d), $w = 4\,w_o$ becomes the most abundant period (blue arrows). In addition, a domain size of $w = 8\,w_o$ (violet arrow) can also be observed. So the transition to the a/c phase takes place by sequentially annihilating one every other b-domain and, thus, the apparent domain periodicities follow a Cantor set sequence with $w = 2^n w_o$.[45] Pre-fractal domain patterns following Cantor set sequences have been observed in ferroelectrics under electric field[46].

Domain structures are also observed in BaTiO$_3$ surfaces during the phase transition at 120°C, where the pattern formation is extended over wide thermal intervals.[47-49] Low energy electron microscopy (LEEM) revealed that the surface near regions in BaTiO$_3$(001) show transient intersections between polar domain walls during the phase transition. Such transient surface structures persist under heating to temperatures above T$_C$ while the bulk has already transformed into the cubic phase. The wall signals are crisscrossed by a second set of stripe patterns with roughly perpendicular orientation at 126.3°C. These surface patterns coarsen under further heating to 126.9°C. Most importantly, the intersections of ferroelastic/ferroelectric 90 degrees walls and the surface are electrically charged.

In our new experiments, branching points are found (white circle Figure 1d) where two a/c domains merge into one, similar to those reported in various works[13,15,37,50]. All these bifurcations have a b-domain passing through them, showing that they are preferred nucleation points for new b-domains.[29] Due to the random distribution of these defects, slightly different domain sizes can be present and kinks (or less straight domain wall) are found for some of the newly formed b-domains (Figure 1e-f), due to the merger of two b-domains with different nucleation points. In these cases, a deviation from the intrinsically favorable $2^n$ domain evolution law. Indeed, we have occasionally observed tripling of domain widths.

Interestingly, these experiments have also shown the self-repairing of "wrongly" nucleated domain walls, such as those signaled by the dashed green markers in Figure 3c. It is then interesting to note that the system is able to "correct" small discrepancies in domain sizes after the domains have already been formed. Figure 3a shows b-domains right after their formation, displaying unequal widths due to the local structure. The elastic forces in this situation lead to a movement of the b-domains with respect to each other to reach their equilibrium position equidistant between the neighboring b-domains (Figure 3b).

Thus, changes of the periodicity take place by the formation of new stress-relieving elements (singularities) exactly in the center of the old domains already predicted by dynamic pattern formation.[51,52] In the previous measurements a new b-domain (thus, a pair of domain walls) conforms such singularity, but other types such as single domain walls, dislocations, cracks, etc. are governed by similar physics. One can simplify such dynamic processes if one considers only one relevant strain component perpendicular to the wall[53], and postulates the strain energy $E \propto \varepsilon^2$, where the strain $\varepsilon$ can be expressed as a function of the domain volume fractions, $\alpha$, as $\varepsilon = (1-\alpha)\varepsilon_1 + \alpha\,\varepsilon_2$. The minimization of the energy $E$ leads to $\alpha = \frac{1}{2}(\varepsilon_1/\varepsilon_2)$. In our case, the two new domains are of the same a-domain type and, thus, $\alpha = \frac{1}{2}$. So the new b-domains will be formed exactly in the middle of the old domain. Increasing the strain (e.g. by decreasing the temperature) would repeat the process between a first-generation singularity and a second generation singularity, and so on. Of course, other strain components may differ. In addition, the walls contain dipoles, which lead to further long-range interactions.[54,55]

Being able to follow the evolution of ferroelastic domains, allows for a better understanding of the properties of the material, which often show signatures of the history of domain formation. One example of this is the domain size modification by pre-existing domain walls that we have observed in our films. As mentioned before, the domains width ($w$) dependence on the film thickness ($d$) is given by[24] $w = \beta\, d^{1/2}$, where the $\beta$ pre-factor varies between systems, as it is primarily a function of the anisotropy of the order parameter.[56,57] Ferromagnetic systems give rise to a larger $\beta$ ($\approx$ 10-100 nm$^{1/2}$) than the $\beta$ found for ferroelectric systems (~1-5 nm$^{1/2}$), for which a strong crystalline anisotropy exists due to the presence of a polar axis.[58] Differences in $\beta$ pre-factors within a single type of material can also be found when the sample morphologies produce different domain wall widths.[59] In the very thin film limit (for $d < w$), Roytburd's approximations are not valid and a rigorous calculation of the elastic energy of the system leads to an approximate linear dependence of $w$ with $d$.[25,26,60,61]

This law, thus, describes the optimal average domain size in thermodynamic equilibrium and in the absence of defects. When kinetic effects affect the final domain configuration, $w = \beta\, d^n$ scaling laws with $½ < n < 1$ have been observed.[32,62]

In our samples, the expected $w = \beta\, d^{1/2}$ is observed (in Figure 4a,b) with $\beta \approx 10$ nm$^{1/2}$ for both the a/b and a/c domains. However, when the a/c domains are obtained upon heating from the a/b ones, $\beta$ is reduced to about 7 nm$^{1/2}$ (Figure 4c). One explanation for this observation is that the domain walls for the new a/c' domains (green-dashed lines in Figure 4d) preferably nucleate at the existing a/b domain walls (black-solid lines). Since these domain walls are rotated by 45° with respect to each other (around the azimuthal axis), the domain width relationship between these two domain structures is governed by a √2 factor. This would explain the approximately same ratio of the observed $\beta$ values for the a/b ($\beta = 10$nm$^{1/2}$) and a/c' ($\beta = 7$nm$^{1/2}$) domain sizes. So pre-existing walls acting as nucleation and pinning sites will still give rise to a quadratic law if they are sufficiently periodic with a modified $\beta$ scaling factor.

From the results of the phase-field simulations, shown in Figure 5, it is easily seen that the alternating pseudo-a/b phase appears at low temperatures and an a/c tetragonal structure predominates at high temperatures. From Figure 5a, it can be seen that the domain structure of the high temperature phase, a/c, agrees with the experimental observations previously captured by Everhardt *et al.* with (101) domain walls. These simulations also demonstrate the change in domain wall orientation from {101} to {110}, consistent with PFM experiments, with the "new" (orthorhombic) domains forming in the middle of the "old" (tetragonal) domains. An analysis of the phase-field simulation results in a $\beta$ coefficient of $\beta = \sim7$ at high temperature and $\beta = \sim10$ at low temperature, supporting our experimental observations.

As the temperature of the system decreases, or increases, the periodicity of the domains halves, or doubles, also in agreement with experimental observations. Investigation of elastic energies agree with analytical theory and experimental observation of stress relaxation and the formation of new domains at the center of old domains. In order to be able to show larger simulation areas, we have also performed 2D simulations. The orientation of the polarization in the two phases is not fully in agreement with the experiment, but successive nucleation of a new domain halfway between existing domain walls is clearly observed as the temperature is decreased. This is consistent with the apparent sequential periodicity halving observed in the PFM experiments. In addition, it is shown that the 'new' domains nucleate near the film-substrate interface, along a 90° (100) domain wall and along the <110> directions.

(CONCLUSIONS)

In conclusion, we directly observe, for the first time, sequential periodicity halving/doubling of ferroelastic/ferroelectric domains through a phase transition in BaTiO$_3$ thin films. This demonstrates that the maximum probability for the nucleation of a new domain is exactly in the middle of an existing domain, halfway between two existing domain walls. We, thus, observe a characteristic transitional sequence: the first generation of domains is periodic, the second generation nucleates in the middle between existing domains, the third again in the middle between second generation domains and so on. Similar sequences were found as part of clock-model calculations.[63] This behavior is part of a class of scaling phenomena known as period-doubling cascades that have been observed in nature in a wide range of physical systems and are mathematically investigated by means of bifurcation theory.[64-67] Despite such phenomena are well-known for dynamical systems and a link with spatially modulated phases of matter has been made[1], they have not yet been observed experimentally. We believe that our observation of domain periodicity halving is not exclusively characteristic of the system under investigation, but represents a more general mechanism of transformation in materials with periodic structures.

FIGURES

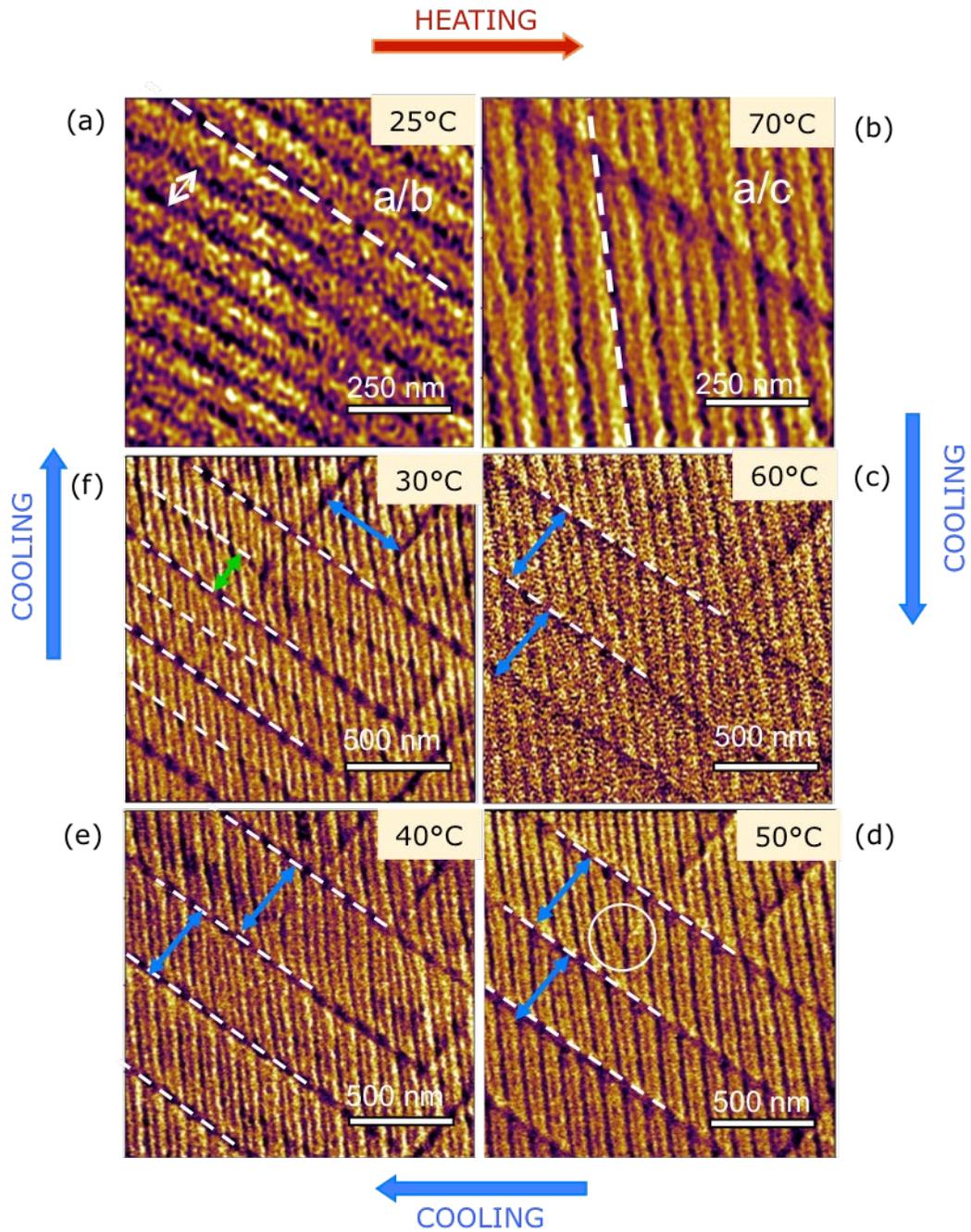

**Figure 1.** Domain evolution during cooling. Amplitude LPFM images (in DART mode) of an 88 nm thick BaTiO$_3$ film on a NdScO$_3$ substrate. (a) shows the domain configuration before the temperature cycling (at 25°C) with a/b domains. The dashed line indicates the direction of the domain walls, parallel to the [110] direction. The white double arrow in (a) signals the domain width at 25°C, which is $w_o \sim$ 110nm. Afterwards the sample was heated to the paraelectric phase at 200 °C for 5 minutes and then cooled down in steps. Measurements were performed at 70 °C (b), 60 °C (c), 50 °C (d), 40 °C (e) and 30 °C (f). In (b) the sample is mainly in the a/c domain configuration, the white dashed line signaling the direction of the a/c domain walls. While cooling down further (c,d,e,f) the a/b domains appear progressively in the center of the existing domains. The blue and green double arrows indicate the observed a/b domain widths of 500 nm and 250 nm, respectively.

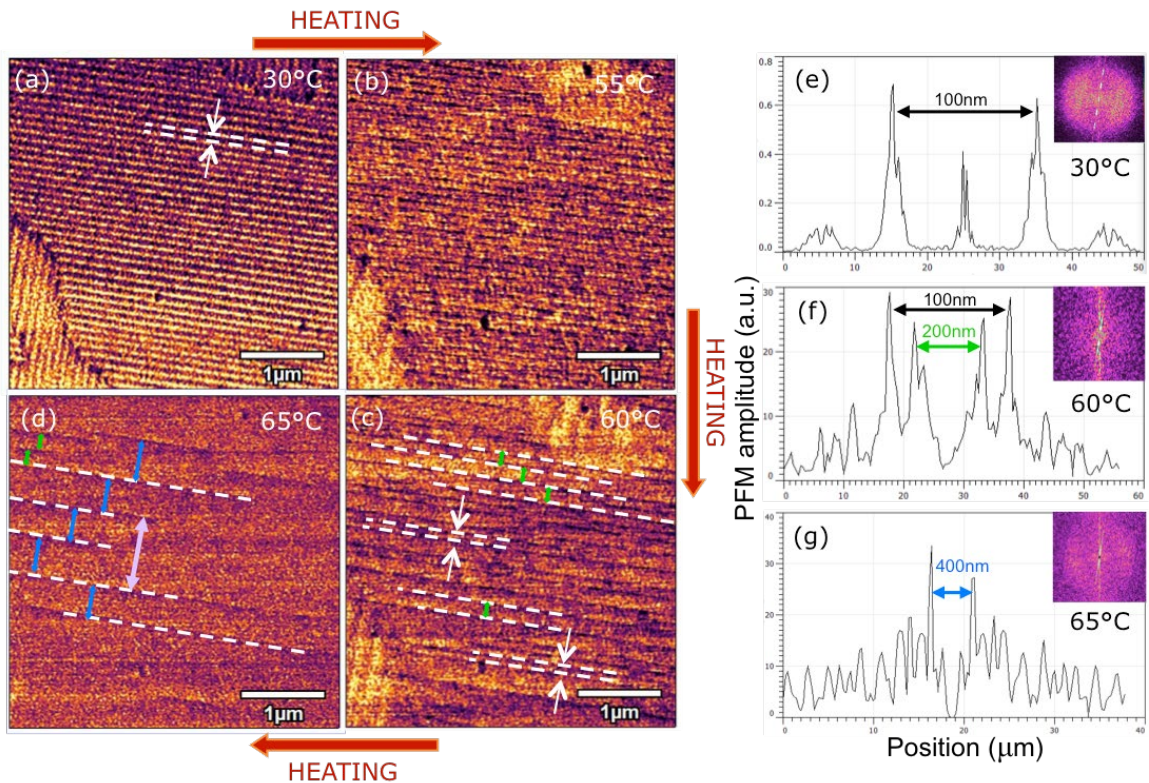

**Figure 2.** Domain evolution during heating. Amplitude of the LPFM signal of a 90 nm thick BaTiO$_3$ film on a NdScO$_3$ substrate measured during heating. The transition from a/b to a/c domains is followed on heating from 30°C to 70°C at 5° steps. Here, the images at 30 °C (a), 55

°C (b), 60 °C (c) and 65 °C (d) are shown. The most prominent domain widths (*w*) are signaled by colored arrows: *w= w₀=* 110 nm (white) in a), b) and c); *w=2w₀* = 200 nm (green) in c) and d) and *w=4w₀* 400 nm (blue) and *w= 8w₀=* 800 nm (violet) in d). The domain period is obtained by the Fast Fourier Transforms (FTT) of the PFM images at 30 ºC (e), 60 ºC (f) and 65 °C (g).

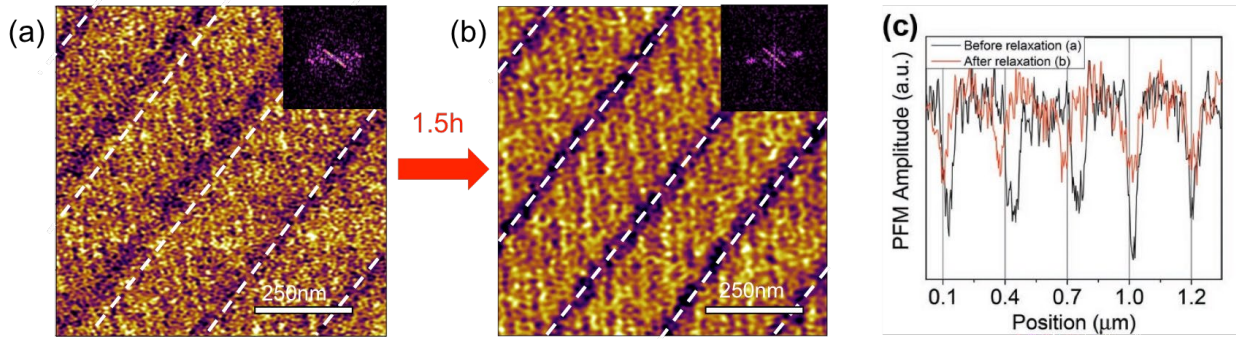

**Figure 3.** Self-repairing periodicity. LPFM amplitude images of a 170 nm thick $BaTiO_3$ film on a $NdScO_3$ substrate. (a) shows five different b-domains (darker diagonal constrast) with slightly different distances between them (different a-domain widths). The dashed lines signal the position of the b-domains in the ideal periodic case (b) After waiting for 1.5 hours, the domain walls move with respect to each other in order to achieve equal distances. The insets in (a,b) are the Fast Fourier Transforms (FTT) showing a more periodic structure in (b). (c) Line scan taken along the diagonals of the PFM amplitude images in (a) and (b). The dips are the positions of the b-domains and it can be seen that the position of some b-domains has changed after the relaxation.

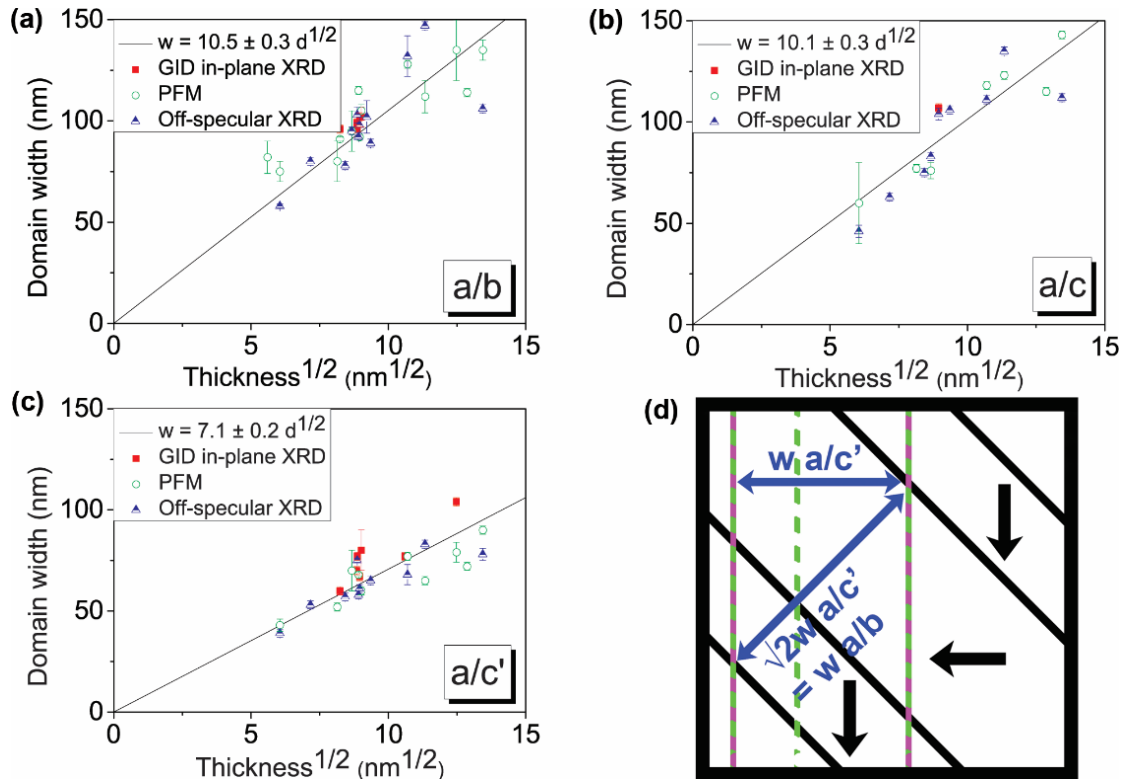

**Figure 4.** Domain size scaling. Domain widths for different film thicknesses, determined by either in-plane Grazing Incidence XRD (red filled squares), off-specular (204) XRD peaks (blue half-filled triangles) or local PFM (green open circles) (methods described in ref.[40]). The domain size is fitted to a Roytburd square root law with parameters is given in the legend for the a/b domain configuration (a), for the a/c state obtained by cooling down from the paraelectric phase (b) and for the a/c' domain configuration obtained by heating from the a/b state (c). (d) Sketch of the possible mechanism to explain the differences between (b) and (c), showing the a/b matrix (black diagonal lines) with the newly formed, upon heating, a/c domain walls (green/purple dashed and green/white dashed lines are the two types of domain walls).

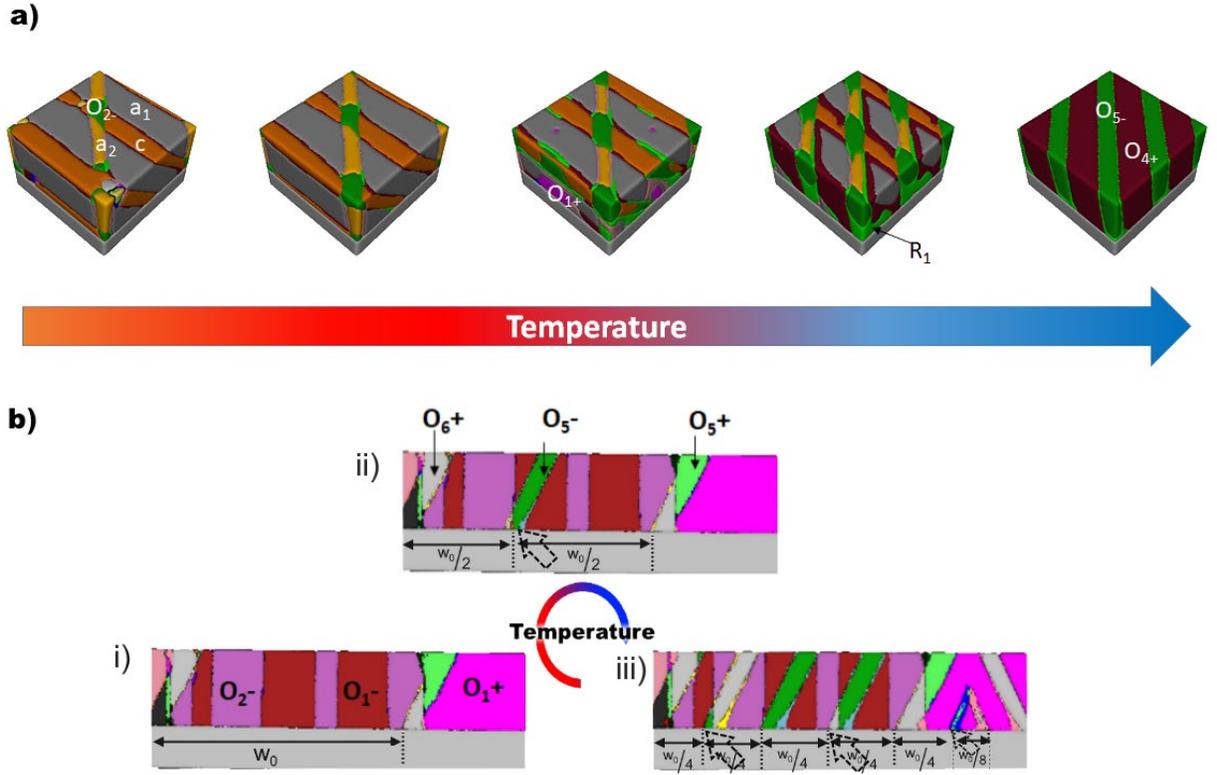

**Figure 5.** Phase Field Simulations of BaTiO$_3$ thin films. (a) Three-dimensional case. The temperature decreases from 100°C to 25°C and the apparent halving phenomena is occurring with the change in domain wall orientation from {101} to {110} across the phase transition. (b) Two-dimensional case. Even though the polar orientations do not fully correspond to those of the experiment, subsequent nucleation of a new domain halfway between existing domain walls (shown by the open arrows) is shown to take place as the temperature is decreased from i) to iii). Domain Definitions: $a_1 = (P_0,0,0)$, $a_2 = (0,P_0,0)$, $c = (0,0,P_0)$, $O_{1+} = (P_0, P_0, 0)$, $O_{1-} = (-P_0, -P_0, 0)$, $O_{2+} = (P_0, -P_0, 0)$, $O_{2-} = (-P_0, P_0, 0)$, $O_{4+} = (P_0, 0, -P_0)$, $O_{5+} = (0, P_0, P_0)$, $O_{5-} = (0, -P_0, -P_0)$, $O_{6+} = (0, P_0, -P_0)$, $O_{6-} = (0, -P_0, P_0)$, and $R_1 = (P_0, -P_0, -P_0)$.

ASSOCIATED CONTENT

**Supporting Information**.

**Note S1**

The presence of different coexisting domain widths in epitaxial thin films is ubiquitous in the literature.[31-37] Careful analysis shows that these widths often appear to be governed by the $2^n$ relationship expected from periodicity halving/doubling. As example we can point to Figures 1a,b) in ref.[32], where the different sizes of c-domains that coexist along the b-axis of a PbTiO$_3$ thin film grown on a DyScO$_3$ substrate are 75, 150 and 300 nm (see Figure S1b). Another example are the PbZr$_{0.2}$Ti$_{0.8}$O$_3$, thin films grown on SrTiO$_3$ by Ganpule et al.[32], which have c-domain sizes of 22, 44, 88 and 172 nm, also related by period doubling (Figure S1a). More examples with similar phenomena can be found in the literature.[31,32,34–37]

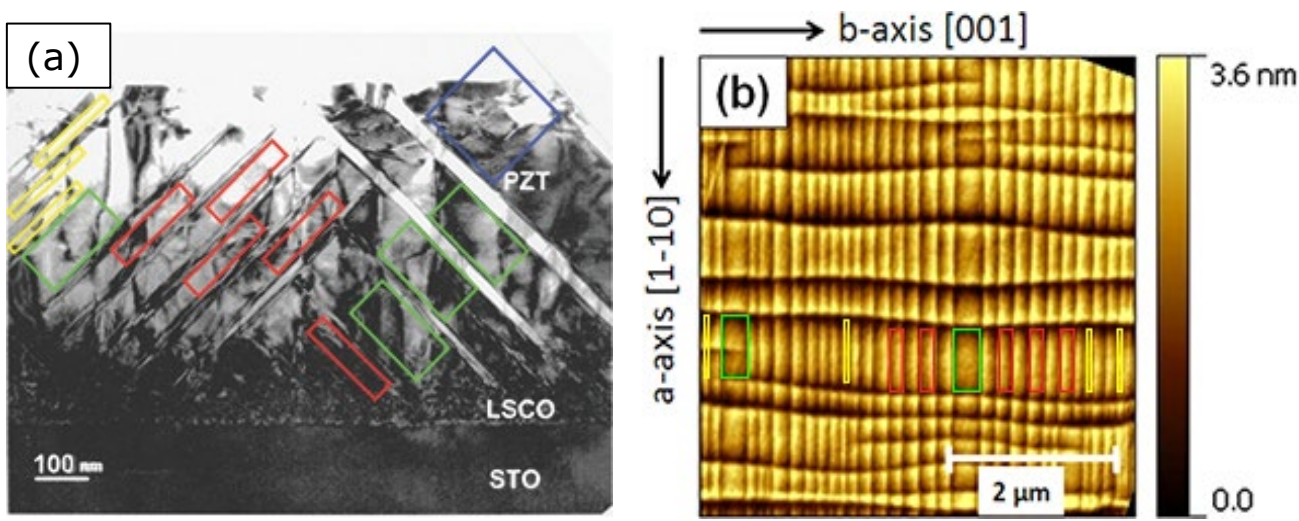

**Figure S1.** (a) Cross-sectional TEM image of PbZr$_{0.2}$Ti$_{0.8}$O$_3$ thin films grown on SrTiO$_3$ by Ganpule et al.[33] Few specific domain sizes (only considering c-domains here) can be found with an error of about 5 nm. The yellow, red, green and blue rectangles highlight the 22 nm wide, 44

nm wide, 88 nm wide and 172 nm wide domains, respectively. (b) PFM image of PbTiO$_3$ films grown on DyScO$_3$ by Nesterov et al.[32]. Only a select few domain sizes (only considering c-domains here) can be found with an error of about 10 nm. The yellow, red and green rectangles indicate some of the 75 nm, 150 nm and 300 nm wide domains, respectively.

**Note S2**

The growth conditions were not identical for all the BaTiO$_3$ thin films available and the quality of the films varied. In order to extract the domain sizes for Figure 4 of the main text, only those films for which all the available methods gave consistent results, were used. Those methods are in-plane Grazing Incidence XRD, off-specular (204) XRD peaks and local PFM. Details about them can be found in ref. [38]. The three techniques can sometimes give quite different results if the local environments does not allow for full relaxation into a/b domains at certain positions, or if a mixture of a/c and a/c' is found because the a/b phase is never fully reached. The domain sizes that have been obtained for all the available films, prior to the above-mentioned selection, is given in Figure S2. It can be seen that the spread of domain sizes around the square root law fit is larger, but the best fit gives the same pre-factor $β$ as that obtained from Figure 4.

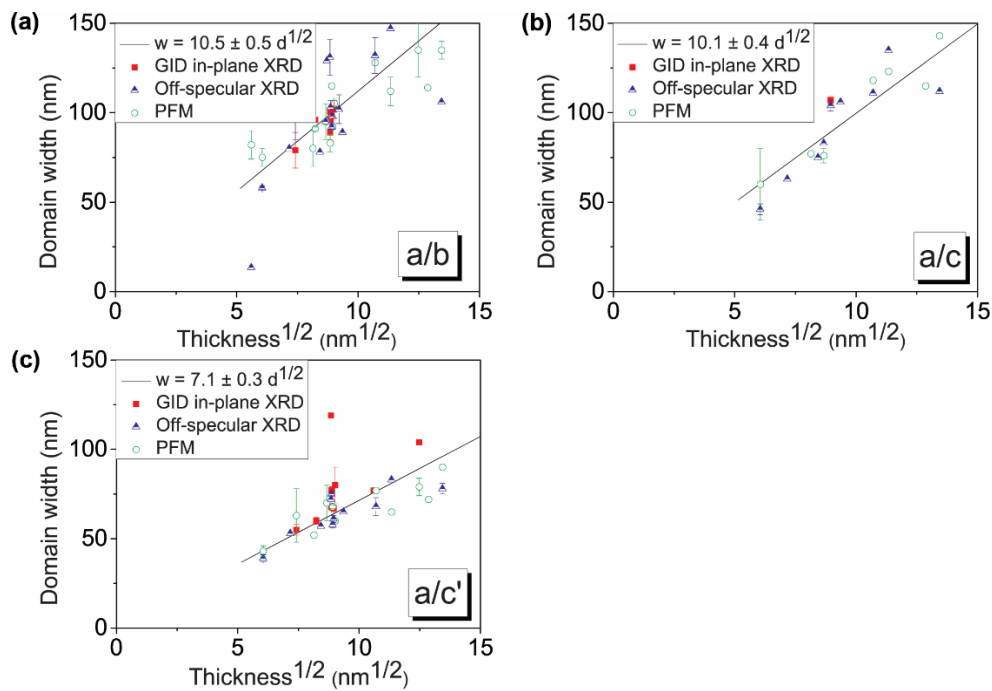

**Figure S2.** The domain size scaling for different film thicknesses using all available samples with variable quality and different experimental methods. The domain size is fitted to a square root law, for the a/b phase (a), for the a/c phase obtained by cooling down from the paraelectric phase (b) and for the a/c' phase obtained by heating from the a/b phase (c).

**Note S3**

Coexistence of a/b and a/c domain structures at room temperature has been observed in some of the BaTiO$_3$ thin films by means of PFM and XRD, as also reported in ref.[40].

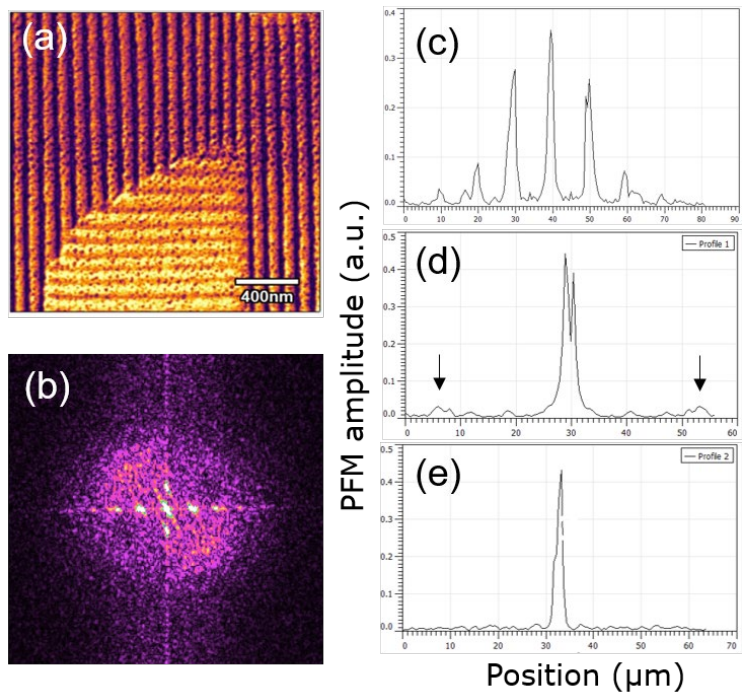

**Figure S3.** Room temperature coexistence of a/b and a/c domains structures. (a) Lateral amplitude PFM image of a (90 nm thick) $BaTiO_3$ film and (b) its FFT, where two 90° a/b superdomains[39] are clearly visible. (c) Periodicity peaks corresponding to the a/b domain structure. (d-e) Profiles of the FFT at 45° from the horizontal direction: in the second and fourth quadrant two small peaks are visible (black arrows) corresponding to the periodicity of one single orientation of the a/c domain structure.

AUTHOR INFORMATION

**Corresponding Authors**

*E-mail: s.damerio@rug.nl; b.noheda@rug.nl

**Present Addresses**

#Presently at: Materials Science Division, Lawrence Berkeley National Laboratory, Berkeley, CA 94720, USA.

ACKNOWLEDGMENT

The authors are grateful to Marty Gregg, Janusz Przeslawski, Jim Scott and Yachin Yvry for useful discussions. A.S.E., S.D. and B.N. acknowledge financial support from the alumni organization of the University of Groningen, De Aduarderking (Ubbo Emmius Fonds), the Zernike Institute for Advanced Materials and the Groningen Cognitive Systems and Materials (CogniGron). Parts of this research were carried out at the light source Petra III at DESY, a member of the Helmholtz Association (HGF). G.C. and N.D. acknowledge ERC Starting Grant 308023 and projects FIS2015-73932-JIN and MAT2016-77100-C2-1-P from the Spanish MINECO and 2017 -SGR-579 project from the Generalitat de Catalunya. All work at ICN2 is also supported by the Severo Ochoa Program (Grant No. SEV-2017-0706). The work at Penn State is supported by the U.S. Department of Energy, Office of Basic Energy Sciences, Division of Materials Sciences and Engineering under Award FG02-07ER46417 (J.A.Z. and L.Q.C.) and partially by a graduate fellowship from the 3M Company (J. A. Z.). Computations for this research were performed on Pennsylvania State University's Institute for CyberScience Advanced CyberInfrastructure (ICS-ACI).

ABBREVIATIONS

XRD, X-Ray Diffraction; PFM, Piezoresponse Force Microscopy; TEM, Transmission Electron Microscopy; DART, Dual AC Resonance Tracking mode; FFT, Fast Fourier Transform.